\documentclass[aps,pra,twocolumn,superscriptaddress]{revtex4-1}

\usepackage{graphicx}
\usepackage{dcolumn}
\usepackage{bm}
\usepackage{hyperref}

\begin{document}


\title{Disorder Enhanced Nuclear Spin Relaxation at Landau Level Filling Factor One}



\author{Tong Guan}
\affiliation{Beijing National Laboratory for Condensed Matter Physics, Institute of Physics, Chinese Academy of Sciences, Beijing 100190, China}

\author{Benedikt Friess}
\affiliation{Max Planck Institute for Solid State Research, Heisenbergstra\ss e 1, D-70569, Stuttgart, Germany}

\author{Yongqing Li}
\email[]{yqli@iphy.ac.cn}
\affiliation{Beijing National Laboratory for Condensed Matter Physics, Institute of Physics, Chinese Academy of Sciences, Beijing 100190, China}

\author{Shishen~Yan}
\affiliation{School of Physics, Shandong University, Jinan 250100, China}

\author{Vladimir Umansky}
\affiliation{Weizmann Institute of Science, Rehovot 76100, Israel}

\author{Klaus von Klitzing}
\affiliation{Max Planck Institute for Solid State Research, Heisenbergstra\ss e 1, D-70569, Stuttgart, Germany}

\author{Jurgen~H.~Smet}
\email[]{j.smet@fkf.mpg.de}
\affiliation{Max Planck Institute for Solid State Research, Heisenbergstra\ss e 1, D-70569, Stuttgart, Germany}


\date{\today}

\begin{abstract}
The nuclear spin relaxation rate (1/$T_1$) is measured for GaAs two-dimensional electron systems in the quantum Hall regime with an all-electrical technique for agitating and probing the nuclear spins. A "tilted plateau" feature is observed near the Landau level filling factor $\nu=1$ in 1/$T_1$ versus $\nu$. Both width and magnitude of the plateau increase with decreasing electron density. At low temperatures, 1/$T_1$ exhibits an Arrhenius temperature dependence within the tilted plateau regime. The extracted energy gaps are up to two orders of magnitude smaller than the corresponding charge transport gaps. These results point to a nontrivial mechanism for the disorder enhanced nuclear spin relaxation, in which microscopic inhomogeneities play a key role for the low energy spin excitations related to skyrmions.
\end{abstract}


\maketitle


\section{Introduction}

Because the relaxation of nuclear spins can  proceed via the hyperfine interaction with electrons, measurement of the nuclear spin relaxation rate provides a powerful means of probing complex and rich spin related electronic structures that exist in various systems. This is particularly true for the two dimensional electron system (2DES) exposed to a perpendicular magnetic field $B$ (for reviews see Refs.~\cite{Li08,Hirayama09}). In case of a sufficiently strong magnetic field, the energy spectrum is discretized and the $B$-field also causes a large mismatch between the Zeeman energies of electrons and  nuclei. It suppresses the nuclear spin relaxation via the hyperfine interaction, since the electron-nuclear spin flip-flop processes are forbidden by the law of energy conservation. Nuclear spin relaxation via the hyperfine interaction is however possible, if the Zeeman energy mismatch can be compensated in some other features in the energy spectrum. Examples include the existence of an overlap between the disorder broadened Landau levels of opposite spins~\cite{Berg90}, differently populated edge channels of opposite spin orientations~\cite{Dixon94,Kawamura07}, the presence of interaction induced gapless low energy excitations involving spin flips as they occur in a Skyrme crystal~\cite{Tycko95}, and the coexistence of domains with opposite spin originating from a spin-related phase transition in the electronic system as a result of disorder~\cite{Smet02,Hashimoto02,Zhang07}.  Nuclear spin relaxation measurements have played an important role in unveiling the nature of complex low energy excitations and spin-related phase transitions in the 2D systems~\cite{Li08,Hirayama09}.

Numerous experiments have reported a  strong enhancement of the nuclear spin relaxation rate near {\em but not at} Landau level filling factor $\nu\equiv nh/(eB)=1$, where $n$ is the 2D electron density, $h$ is the Planck constant, $e$ is electron charge, and $B$ is perpendicular magnetic flux density (i.e. $\mathbf{B}$-field component perpendicular to the 2DES plane)~\cite{Tycko95,Smet02,Desrat02,Hashimoto02,Gervais05,Tracy06,Bowers10}.  When the strength of disorder is weak, the ground state at $\nu=1$ is a fully spin-polarized quantum Hall ferromagnet~\cite{Girvin00}. When the filling deviates from exact $\nu=1$, skyrmions, vortex-like spin textures possessing topological charge, emerge as a consequence of lowering the exchange energy between the electrons~\cite{Skyrme62,Sondhi93,Brey95,Barrett95,Schmeller95,Aifer96}.  In the presence of a sufficient number of skyrmions the nuclear spin relaxation rate (1/$T_1$) is enhanced. This enhancement has frequently been attributed to a gapless Goldstone mode of a Skyrme crystal that forms at low temperatures~\cite{Cote97,Gallais08,ZhuH10}. Even though such low energy excitations of skyrmions provide a natural explanation of the enhanced nuclear spin relaxation, controversy remains in view of discrepancies in the observed temperature dependence of 1/$T_1$~\cite{Tycko95,Gervais05,Tracy06,Bowers10} and in the predicted temperature for the melting of the Skyrme crystal~\cite{Green96,Cote97,Timm98,Green00}. It has also been pointed out theoretically that quantum fluctuations may lead to a liquid ground state near $\nu=1$ at zero temperature~\cite{Paredes99}. When disorder effects are included, the fully spin-polarized quantum Hall ferromagnet from which it all starts may be an oversimplified description of the $\nu=1$ ground state even at zero temperature, and other electronic phases, such as paramagnetic states, partially polarized ferromagnetic states and quantum Hall spin glasses, have been predicted~\cite{Sinova00,Green00,Murthy01,Rapsch02,Makogon10}.  Previous experiments on nuclear spin relaxation at or near $\nu=1$ have, however, not been performed such that the disorder effects on the electron spin structures can be explored in detail, in particular at temperatures below 40\,mK.

In this work, we shed more light on the nuclear spin relaxation issue near and at $\nu=1$ by extending existing studies to samples with different levels of disorder and to temperatures less than 15\,mK. This temperature regime has become accessible by using \textit{all-electrical} nuclear spin relaxation experiments. The role of disorder is studied either by varying the electron density with a back gate or by comparing samples of different quality. The nuclear spin relaxation rate exhibits a tilted plateau structure around $\nu = 1$. This plateau vanishes with improved sample quality indicating an important role of disorder. Previously invoked mechanisms~\cite{Berg90,Iordanskii91,Antoniou91} to account for disorder enhanced nuclear spin relaxation near odd filling factors are not compatible with our data.

\section{Experiments}

\begin{figure*}
\centering
\includegraphics*[width=12 cm]{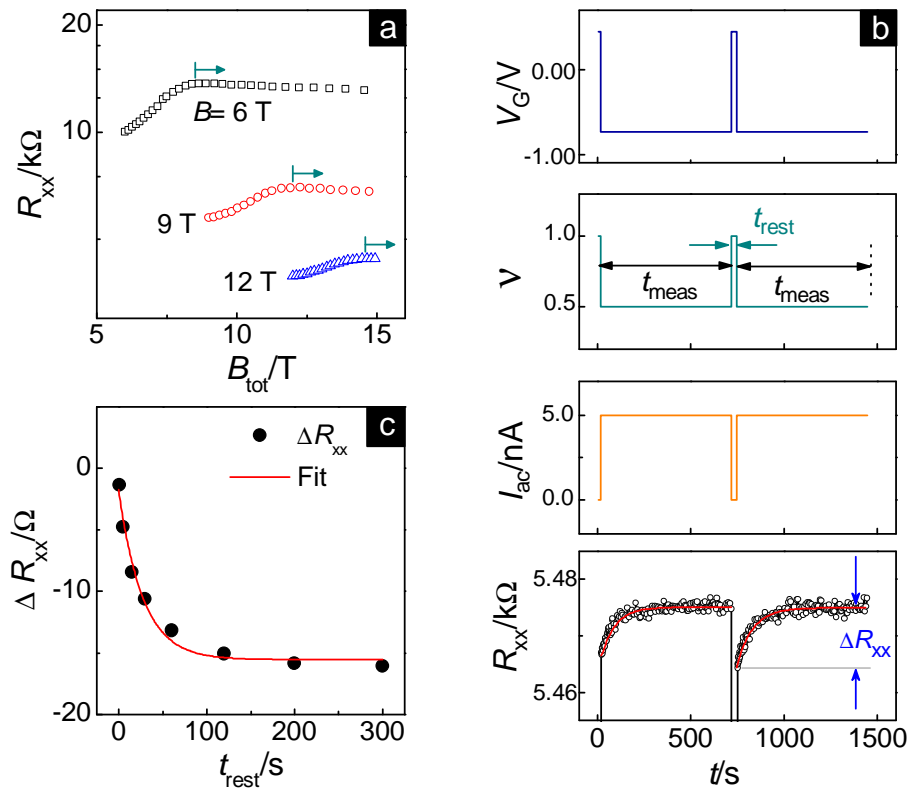}  
\caption{\label{Fig1} (a) $R_\mathrm{xx}$ at $\nu=1/2$ as a function of the total magnetic field, $B_\mathrm{tot}$, at the base temperature for perpendicular fields $B=$ 6, 9, and 12\,T. Increasing of $B_\mathrm{tot}$ leads to transitions from partial to full spin polarization, which are marked by short vertical lines. The arrows denotes regions of full spin polarization. (b) A fraction of the measurement sequence for the nuclear spin relaxation measurements. The bottom panel shows an example of $R_\mathrm{xx}$-$t$ trace (circles) and the exponential decay fit to Eq.\,(1) (line). (c) An example of how to extract time constant $\tau_1$ with a fit of the $\Delta R_\mathrm{xx}$ data extracted from the procedure illustrated in panel (b) to Eq.\,(2).}
\end{figure*}

Our studies reported here were mainly performed on two GaAs/AlGaAs quantum well wafers. Both wafers were patterned into 400\,$\mu$m Hall bar shaped samples equipped with a backgate. In sample A, the 2DES resides in a 16\,nm wide quantum well and the electron density $n$ varies with the backgate voltage $V_G$ in units of Volts according to $n=(1.79+0.89V_G)\times10^{11}$\,cm$^{-2}$. Sample B has a 30\,nm wide quantum well and $n=(1.79+0.69 V_G)\times10^{11}$\,cm$^{-2}$. The electron mobilities at $V_G=0$ are $0.8\times10^6$ and $1.3 \times10^7$\,cm$^{2}$/Vs for samples A and B, respectively. The data shown below were taken on sample A unless otherwise specified.

The all-electrical method to detect the nuclear spin relaxation rate at an arbitrary filling factor $\nu_{\rm rest}$ only requires quasi-dc electron transport~\cite{Li12}. The nuclear spin system is first driven out of equilibrium to elicit a time-dependent response, for instance by imposing an electrical current through the sample. Subsequently, the recovery to equilibrium is monitored. At the $B$-fields relevant here, the Fermi sea of composite fermions at $\nu=1/2$ is partially polarized. The two spin populations have a continuous and finite density of states. This enables Korringa-like hyperfine coupling, so that the nuclear spin system can efficiently interact with the electronic spin system~\cite{Li08}. This property is exploited to drive the nuclear spin system out of equilibrium. The nuclear spin polarization is estimated to be approximately $10\%$ if the system is left to equilibrate sufficiently long at base temperature. A small ac current is imposed in order to heat up the electronic system. Entropy is transferred via the hyperfine coupling and the existing thermal nuclear spin polarization is partially annihilated.  The longitudinal resistance $R_\mathrm{xx}$ at $\nu=1/2$ depends on the electron Zeeman energy as long as the composite fermion sea is partially spin polarized. This is demonstrated  in Fig.~1a where $R_{\rm xx}$ at $\nu = 1/2$ is plotted as a function of the total magnetic field $B_{\rm tot}$ for three different values of the perpendicular $B$-field. The value of $R_{\rm xx}$ increases with increasing total field, i.e.~Zeeman energy, and then saturates when the composite fermion sea is fully spin polarized (marked by arrows in Fig.~1a). A detailed description of the observed behavior can be found in Refs.~\cite{Tracy07,Li09}. The degree of nuclear spin polarization modifies the composite fermion Zeeman energy. Therefore, changes in the nuclear spin polarization can be monitored by recording the evolution of $R_\mathrm{xx}$ at $\nu=1/2$.

The measurement sequence to detect the nuclear spin relaxation rate at filling $\nu_{\rm rest}$ is illustrated in Fig.~1b. Initially, a small low frequency ac current at $\nu=1/2$ is imposed to increase the electron temperature and partially depolarize the nuclear spin system~\cite{Li12}. Simultaneously, $R_{\rm xx}$ at $\nu=1/2$ can be measured. The current is switched off and the gate voltage is then set to reach $\nu_{\rm rest}$ where the system is left for some time $t_\mathrm{rest}$, during which the nuclear spin polarization relaxes towards the equilibrium value for $\nu_\mathrm{rest}$. Subsequently, the filling is set back to 1/2 for a sufficiently long time ($t_\mathrm{meas}$). The ac current is turned on as before in order to measure $R_\mathrm{xx}$ and partially depolarize the nuclear spin system to prepare for the next sequence. As illustrated in Fig.\,2b, the trace of $R_\mathrm{xx}$ at $\nu=1/2$ vs.\ $t$ is fitted to the exponential decay function:
\begin{equation}
R_\mathrm{xx}(t)=R_0+\Delta R_\mathrm{xx} \exp(-t/\tau).
\end{equation}
Note that only $\Delta R_\mathrm{xx}$ is of interest, since $1/\tau$ does not reflect the nuclear spin relaxation rate for $\nu_\mathrm{rest}$ but for $\nu = 1/2$. In the examples of Fig.~1b, $\Delta R_{\rm xx} < 0$. The measurement sequence is repeated for different $t_\mathrm{rest}$ and $\nu_\mathrm{rest}$. The ($\Delta R_\mathrm{xx}$, $t_\mathrm{rest}$) data is then fitted to a similar exponential function:
\begin{equation}
\Delta R_\mathrm{xx}(t_\mathrm{rest})=\Delta R_\infty+\Delta R\exp(-t_\mathrm{rest}/\tau_1).
\end{equation}
The extracted time constant $\tau_1$ yields the relaxation rate at $\nu_\mathrm{rest}$ (Fig.~1c). If $\Delta R_\mathrm{xx}$ is not too large, $\tau_1$ is close to the nuclear spin relaxation time $T_1$~\cite{Li08}. If not, the required correction can be easily carried out as described in Ref.~\cite{Li12}. This all-electrical method is not limited to the use of $\nu=1/2$. Also, the spin transition at $\nu\approx2/3$ can be similarly used for the all-electrical spin relaxation measurements~\cite{Hashimoto02,Smet04,Li12}.

The nuclear spin relaxation measurements reported here were carried out in two top-loading dilution refrigerators with superconducting magnets. Extensive low pass filtering in the electrical wirings have been implemented in order to obtain low electron temperatures. The temperature was monitored with calibrated ruthenium oxide sensors mounted near the location of the samples. Measurements of fractional quantum Hall states with small activation gaps (such as $\nu=5/2$ states~\cite{Nuebler10}) were used to confirm the electrons in the quantum Hall regime can be cooled down to at least 15\,mK in both dilution refrigerator systems.

\section{Results and discussion}

\begin{figure*}
\centering
\includegraphics*[width=12 cm]{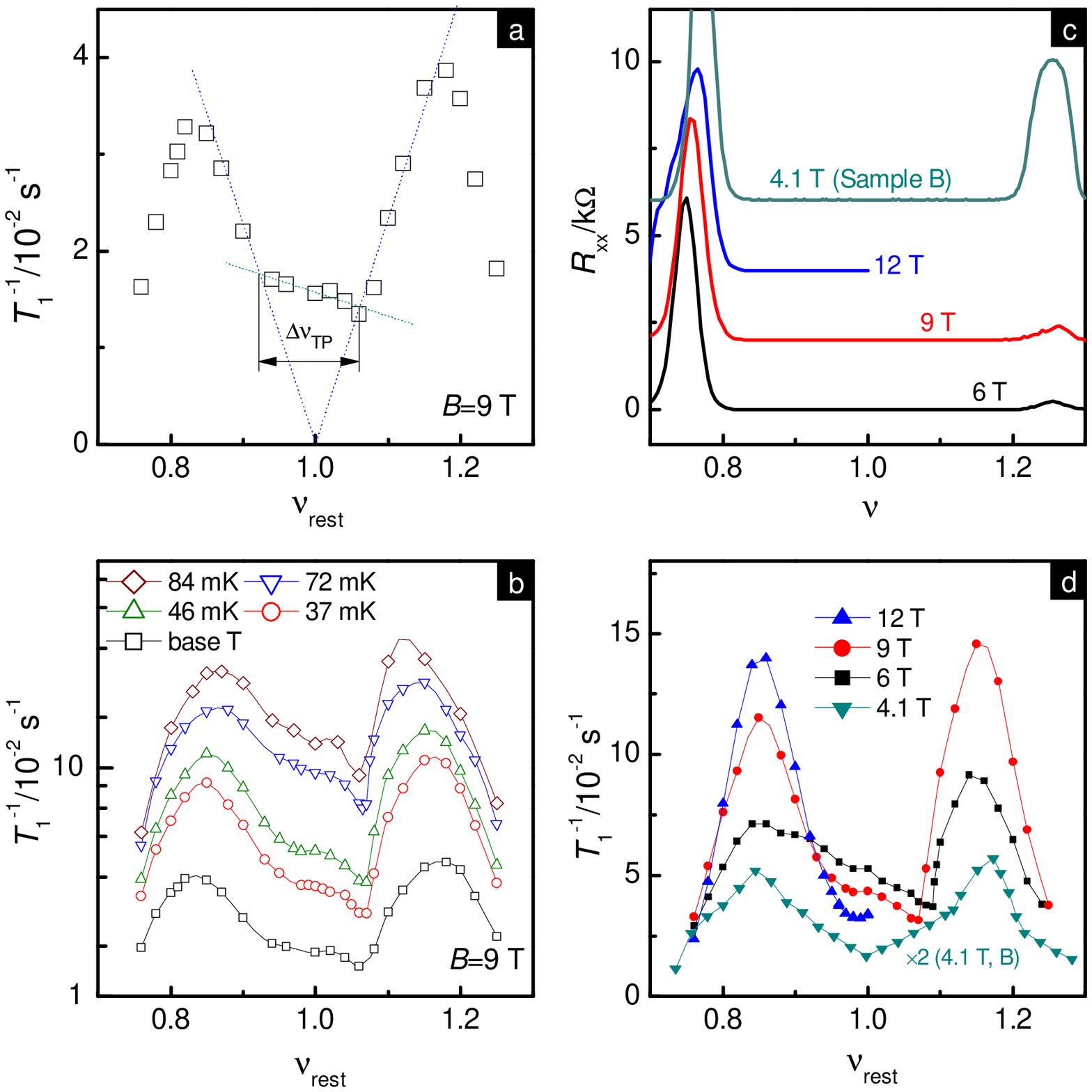}  
\caption{\label{Fig2} (a) Filling factor dependence of the nuclear spin relaxation rate, $1/T_1$, measured at $B=9$\,T at the base temperature for sample A. The experimental data (symbols) are compared with the model of C\^{o}t\'{e} et al.~\cite{Cote97} proposed for disorder-free Skyrme crystals (dotted lines following $1/T_1\propto |\nu-1|$). The width of the tilted plateau is labeled as $\Delta\nu_\mathrm{TP}$.  (b) Filling factor dependences of $1/T_1$ for sample A with $B=9$\,T at $T=$ 10, 37, 42, 72, and 84\,mK (from bottom to top). (c-d) Filling factor dependences of the longitudinal resistance $R_\mathrm{xx}$ (c) and $1/T_1$ (d) at $B=6$, 9, 12 \,T for sample A and 4.1\,T for sample B. The measurements were carried out at $T=$46 mK and the base temperature for samples A and B, respectively. In panel (c), the $R_\mathrm{xx}$-$\nu$ curves are offset by $2\ {\rm k\Omega}$ intervals for clarity.}
\end{figure*}

Fig.~2a displays the filling factor dependence of the relaxation rate measured with $B=9$\,T at the base temperature.
 The 1/$T_1$ data exhibit pronounced peaks at $|\nu-1|\approx0.18$. These peaks are qualitatively similar to those observed previously by Barrett et al.\ with optically pumped NMR at 2.1\,K~\cite{Tycko95} and by resistive detection at $\nu\approx2/3$~\cite{Hashimoto02,Smet02}. They were attributed to the gapless Goldstone mode of the Skyrme crystal. According to the theory of C\^{o}t\'{e} et al., the enhancement of the nuclear spin relaxation rate is proportional to both $|\nu-1|$ and the temperature~\cite{Cote97}. We will comment on this later.
  The drop of $1/T_1$ beyond the peaks (i.e. $|\nu-1|> 0.18$) may just reflect that skyrmionic excitations are no longer favorable at high densities (larger $|\nu-1|$) and the system may give way to quantum liquids as $|\nu-1|$ increases~\cite{Paredes99}.

An unusual feature of the 1/$T_1$ vs.~$\nu_\mathrm{rest}$ data is the "tilted plateau" structure, which exists between the two 1/$T_1$ peaks and around $\nu=1$. As shown in Fig.\,2a, $1/T_1$ could be much larger than that expected for a perfect Skyrme crystal within the plateau. Such a structure is robust against increasing temperature. Fig.\,2b shows that similar plateau shows up at temperatures up to 84\,mK for $B=9$\,T despite much larger values of $1/T_1$ at higher $T$. Similar measurements were also carried out with different magenetic fields for sample A and also for sample B. For the sake of identifying features, $R_\mathrm{xx}$ is shown in Fig.~2c as a function of the filling factor for $B=$ 6, 9, 12\,T for sample A and 4.1\,T for sample B. The corresponding filing factor dependences of $1/T_1$ are depicted in Fig.\,2d, which shows that the width of the tilted plateau ($\Delta\nu_{\rm TP}$) decreases significantly as $B$ is increased from 6\,T to 12\,T. At exact $\nu_{\rm rest} = 1$, the nuclear spin relaxation rate is unexpectedly larger than that expected for a perfect quantum Hall ferromagnet~\cite{Cote97}.

For a homogeneous disorder free quantum Hall ferromagnet at $\nu=1$, the lowest energy charged excitations involving electronic spin flips are skyrmions, and the lowest energy neutral excitations are spin waves or magnons~\cite{Girvin00}. Both types of excitations require an energy that exceeds by orders of magnitude the energy involved in flipping a nuclear spin~\cite{Schmeller95,Gallais08}. Nuclear spin relaxation assisted by the hyperfine coupling should therefore be strongly suppressed and $1/T_1$ is expected to be vanishingly small at the temperatures encountered in this work. Contrary to these expectations, the observed $1/T_1$ at $\nu_{\rm rest} = 1$ in Fig.~2 on sample A is anomalously large. The nuclear spin relaxation rate at $\nu_{\rm rest} = 1$ is largest at low fields and drops at higher fields as the tilted plateau becomes narrower and less pronounced. On sample B with a much higher mobility the tilted plateau has vanished all together. Based on these observations, we conjecture that the tilted plateau structure is associated with the level of disorder. As we tune the density in sample A to higher values, the bare disorder potential, i.e.~the amplitude of the density variations $\Delta n$, remains approximately the same~\cite{Ilani04}, but the spatial variations in the filling factor become smaller as $\nu$ varies with $\Delta n/n$.

In fact, disorder enhanced nuclear spin relaxation was studied long ago~\cite{Berg90,Iordanskii91,Antoniou91}. In the experimental work by Berg et al.~\cite{Berg90}, $1/T_1$ was found to be correlated with $\sigma_\mathrm{xx}$ (or $R_\mathrm{xx}$) near $\nu=3$ at $T=1.3$\,K, and the data could be fairly well explained by considering the exchanged enhanced Zeeman splitting and disorder induced Landau level broadening. A non-vanishing overlap between the densities of states for the  spin-up and spin-down electronic states allows for Korringa-like nuclear spin relaxation. This simple, phenomenological model, however, does not fit our data. As shown in Fig.~2, the filling factor dependence of $1/T_1$ is uncorrelated with the dependence of $R_\mathrm{xx}$. Fig.~3a illustrates that $1/T_1$ at $\nu=1$ is significantly higher than at $\nu=0.76$ for $T>50$\,mK. In contrast, $\sigma_\mathrm{xx}$, which can be calculated from $R_\mathrm{xx}$ in Fig.\,3b is vanishingly small at $\nu=1$ due to the large thermal activation gap (about 50\,K), whereas $\sigma_\mathrm{xx}$ is quite large at $\nu=0.76$. At $\nu=0.85$, where the nuclear spins relax most rapidly at $T>30$\,mK, $\sigma_\mathrm{xx}$ is also nearly zero (See Figs.~3b and 4a). Hence, the disorder model of Berg et al.\ fails to explain the anomalously large relaxation rate at $\nu=1$ as well as the tilted plateau feature.

We note that not only these relaxation rate measurements shown above but also previous measurements of the filling factor dependence of the {\em electron spin polarization}~\cite{Aifer96,Khandelwal01,Melinte01,Zhitomirsky03} exhibited tilted plateau features around $\nu=1$. For a disorder free quantum Hall ferromagnet, a quick depolarization from full spin polarization is expected as one moves away from exact filling factor 1, i.e.\ from exact matching between the number of flux quanta and that of  electrons. For every unmatched flux quantum or electron, an anti-skyrmion or skyrmion is generated involving more than one electron spin flips~\cite{Sondhi93}. The measured electron spin polarization at $\nu = 1$ is, however, far away from the expected full polarization and only outside of the tilted plateau, the electron spin polarization follows the quick depolarization~\cite{Aifer96,Melinte01,Zhitomirsky03}. Local compressibility measurements with a single electron transistor have revealed that electrons attempt to screen, i.e.\ flatten,  the bare disorder potential by setting up a spatially dependent density profile $\Delta n(x,y)$~\cite{Ilani04}. However, as one approaches exact integer quantum Hall fillings, electrons fail to screen the bare disorder potential in some parts of the sample because the local density of states has been exhausted. As a result, a landscape of anti-dots or dots surrounded by an incompressible lake with the nearby integer filling emerges. If disorder is sufficiently strong, dots and anti-dots may coexist. In these anti-dot regions, electrons are missing and hence anti-skyrmions form. In dot regions, skyrmions are present. These skyrmion and anti-skyrmion puddles will significantly lower the average electron spin polarization. They are bound to be responsible for the observed reduced spin polarization~\cite{Aifer96,Melinte01,Zhitomirsky03}. As we deviate from average filling factor 1, anti-dots shrink, while dots expand or vice versa. This compensation effect may account for the tilted plateau. This puddle picture of skyrmions and anti-skyrmions may also play an important role for the observed plateau in the nuclear spin relaxation rate in view of the close resemblance with the behavior of the electron spin polarization. With increased $B$-field, the same landscape of dots and anti-dots recurs at higher average density~\cite{Ilani04}. Since the bare disorder potential does not vary with the field, the spatial density profile $\delta n(x,y)$ set up by the electrons remains the same. As a consequence, the filling factor variation ($\Delta\nu\sim\Delta n/n$) drops with increasing $B$-field. The tilted plateau width, $\Delta\nu_{\rm TP}$, which likely reflects the spatial variation in filling factor, will therefore shrink with increasing  magnetic field, in qualitative agreement with the experimental data in Fig.~2d.

The temperature dependence of $1/T_1$ may also shed light on the underlying relaxation mechanisms. At $\nu = 1$, the nuclear spin relaxation rate in Fig.~3a exhibits thermally activated behavior within the investigated temperature range. An activation gap $\Delta_s \approx 0.11\ \mathrm{K}$ is extracted from an Arrhenius fit to the data. Apparently an energy barrier prevents the nuclear spin relaxation in the limit of
zero temperature. It is tempting to argue that the incompressible region separating the skyrmion and anti-skyrmion puddles is the origin of this energy barrier. Yet, a microscopic picture is lacking for what kind of low energy electronic spin excitations would emerge in such a disordered spin system and how these excitations would enhance nuclear spin flips while propagating between these puddles.

\begin{figure}
\includegraphics*[width=7 cm]{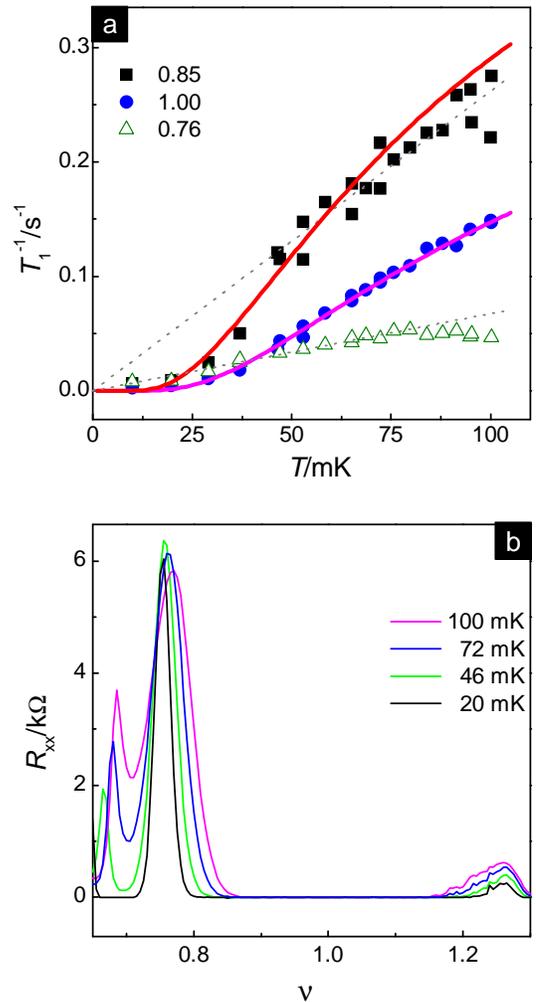}  
\caption{\label{Fig3} (a) Temperature dependence of the nuclear spin relaxation rate $1/T_1$ at $\nu=$ 0.85 (squares), 1.00 (circles), and 0.76 (triangles) with $B=9$\,T. The solid lines are Arrhenius fits to the data. The gray dotted lines illustrates linear temperature dependence expected from the Korringa law. (b) Filling factor dependence of $R_\mathrm{xx}$ at four temperatures between 20 and 100\,mK.}
\end{figure}

Fig.~3a also contains the temperature dependence of $1/T_1$ at other filling factors. It is noteworthy that near $\nu = 0.85$, the relaxation rate is close to the largest among the filling factors investigated in this work and in several previously reported theoretical and experimental studies~\cite{Gervais05,Tracy06,Bowers10,Cote97,Green00}.
The nuclear spin relaxation rate varies approximately linearly with temperature down to about 45\,mK. Below this temperature, $1/T_1$ drops more rapidly in a non-linear fashion. The Arrhenius fit yields an energy gap $\Delta_s \approx 0.08\,{\mathrm K}$, while the gap for charge transport at this filling factor equals $\Delta_c \approx 0.54\,{\mathrm K}$ (Fig.\,4a). A summary of the charge transport gap $\Delta_c$ and the thermal activation energy $\Delta_s$ for the nuclear spin relaxation is displayed in Fig.~4b. For the sake of completeness, we point  out that the linear temperature dependence observed at $\nu = 0.85$ down to 45\,mK is consistent with previous resistively detected NMR data by Tracy et al.~\cite{Tracy06}. Lower temperatures were not accessible in their nuclear magnetic resonance (NMR)-based experiment.  Resistively detected NMR data in Ref.~\cite{Gervais05} suggested, however, an opposite behavior, namely increasing nuclear spin relaxation rate with decreasing temperature.

\begin{figure}
\includegraphics*[width=7 cm]{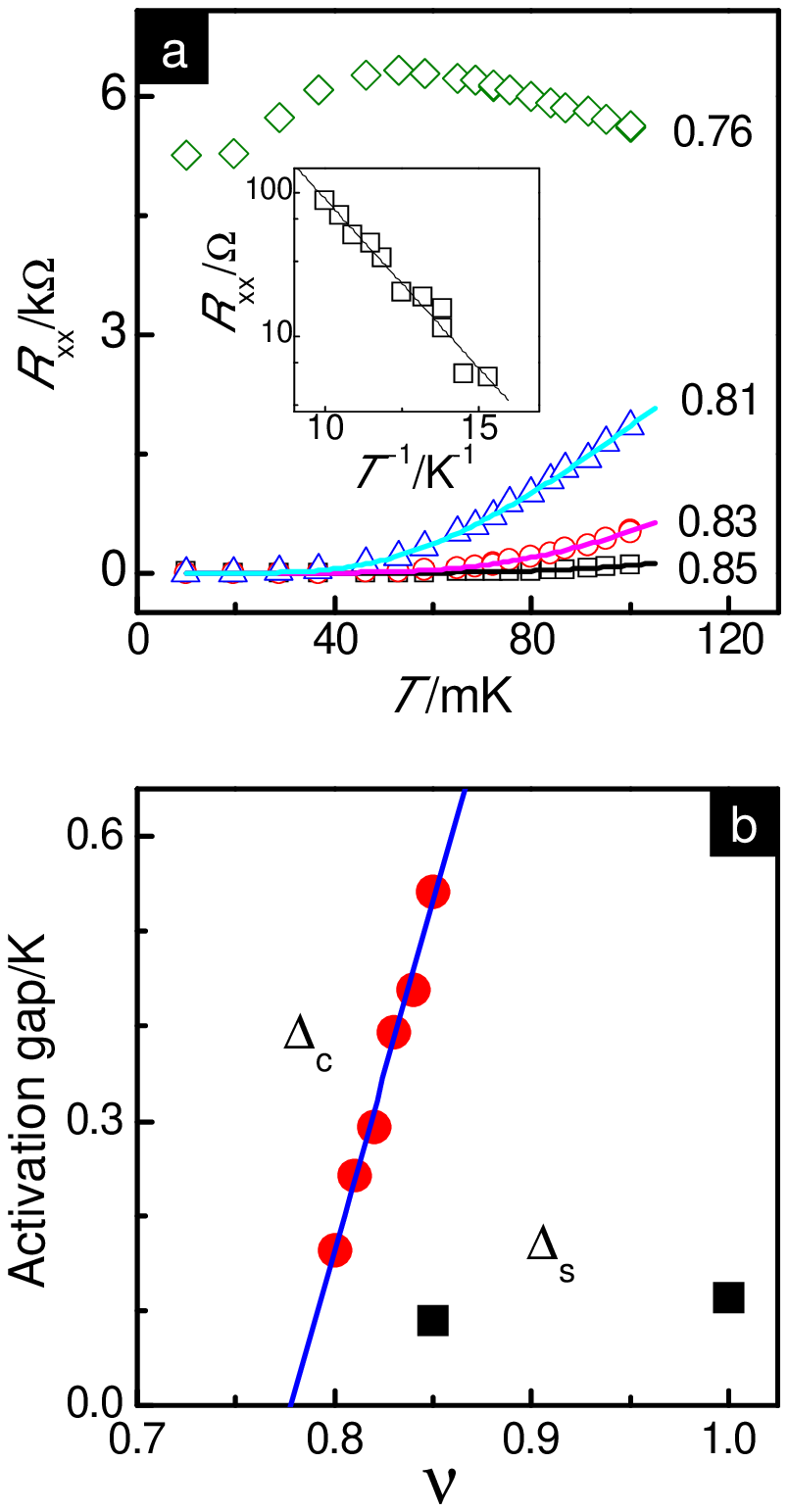}  
\caption{\label{Fig4} (a) Temperature dependence of $R_\mathrm{xx}$ at $\nu=$ 0.76 (diamonds), 0.81 (triangles), 0.83 (circles), and 0.85 (squares) and their fits to the Arrhenius law (lines) at $\nu$=0.81-0.85.  The inset shows the $\log R_\mathrm{xx}$ vs. $1/T$ plot at $\nu=0.85$. (b) Thermal activation gaps $\Delta_c$ (circles) and $\Delta_s$ (squares), at different filling factors, extracted from the Arrhenius fits of the $T$-dependences of $R_\mathrm{xx}$ and $1/T_1$, respectively.}
\end{figure}

C\^{o}t\'{e} et al.~\cite{Cote97} have argued that by moving away sufficiently from $\nu = 1$ skyrmions may form a crystalline lattice, and this lattice possesses a  gapless Goldstone magnon mode that couples strongly to the nuclear spins because it is gapless. It provides a channel for rapid relaxation of nuclear spins. It would give rise to a strongly enhanced relaxation rate with a Korringa-like, i.e. linear $T$-dependence of $1/T_1$. Even when the skyrmionic crystal melts into a liquid state, the Goldstone mode would presumably still be present as an overdamped mode. While the linear temperature dependence above 45\,mK observed here and in Ref.~\cite{Tracy06} is consistent with this picture, the strongly suppressed relaxation at lower temperature is not. It should be noted that disorder is absent in the considerations of Ref.~\cite{Cote97}. Moreover, Bayot et al.\ observed a huge anomalous heat capacity peak at $T\approx40$\,mK in a multiple-quantum-well sample~\cite{Bayot96,Bayot97}. They attributed it to the liquid-to-solid transition of the Skyrme system near $T=40$\,mK. Within the model of Ref.~\cite{Cote97}, the suppressed spin relaxation observed in this work, however, does not favor a transition from a liquid to a crystal as $T$ drops below 40\,mK.

Even though the microscopic picture presented above can capture the main features of our experimental observations, it is still at a crude, qualitative level. Theoretical effort must be taken on the low energy spin excitations emerging from the disordered skyrmion systems in order to fully understand the related spin phenomena. Unfortunately, theories based on Hartree-Fock calculations~\cite{Sinova00,Murthy01}
or more rigorous quantum mechanical treatments~\cite{Rapsch02,Makogon10} have not considered the filling factor and temperature dependences of $1/T_1$. Nevertheless, these theories and a recent experiment on charge transport near $\nu=1$~\cite{PanW11} point to
different ground states (e.g.\ quantum Hall spin glass) other than the simple quantum Hall ferromagnet when disorder is sufficiently strong in comparison to the electron-electron interactions.

\section{Conclusion}
In summary, we have shown that the nuclear spin relaxation measurements can be extended to temperatures well below 40\,mK by using the electrical methods for agitating and probing nuclear spins. The nuclear spin relaxation is found to be enhanced by disorder and exhibits thermally activated behavior. The energy gaps extracted from the nuclear spin relaxation measurements are 1-2 orders magnitude smaller than those from the charge transport. We conclude that none of existing models can quantitatively describe the temperature dependence and the filling factor dependence of the nuclear spin relaxation rate observed in this work. These features point to a nontrivial ground state and low energy spin excitations near filling factor $\nu=1$.

\addcontentsline{toc}{chapter}{Acknowledgment}
\section*{Acknowledgment}
We are grateful to Shahal Ilani, Xin Wan, Kun Yang and Guang-Ming Zhang for valuable discussions.  We acknowledge technical assistance from Wenmin Yang. We acknowledge technical assistance from Wenmin Yang and financial support from the National Basic Research Program of China , the National Science Foundation of China, the Chinese Academy of Sciences, the BMBF, and the German-Israeli Foundation.

\end{document}